# Simulation approach to reach the SQ limit in CIGS-based dual-heterojunction solar cell

Shaikh Khaled Mostaque, Bipanko Kumar Mondal and Jaker Hossain*

*Solar Energy Laboratory, Department of Electrical and Electronic Engineering, University of Rajshahi, Rajshahi 6205, Bangladesh.*

## Abstract

In this article, we demonstrate the design and simulation of a highly-efficient *n*-CdS/*p*-CIGS/$p^+$-CGS dual heterojunction solar cell. The simulation was performed using SCAPS-1D software with reported experimental physical parameters. The simulation performance of our proposed design arises 47% with $V_{oc}$=0.98 V, $J_{sc}$=59.94 mA/cm$^2$ and FF=80.07%, respectively. The high short circuit current and hence the high efficiency is predominantly originated from the longer wavelength absorption of photon through a tail-states-assisted two-step upconversion in dual heterojunction (DH) and thus reaches the SQ detailed balance limit of DH solar cell.

## 1. Introduction

This era is facing a huge demand of electricity, a significant portion of which is being fulfilled by earth's energy sources, especially fossil fuels, resulting pollution and greenhouse effect. A lot of researches are going on for finding alternatives and with natural availability of sun in almost all countries in the world, solar photovoltaics has been found as a prominent, cost effective and eco-friendly way to become an alternative for generation of energy.

Recently, most of the photovoltaic explorations are being focused on obtaining high efficient devices at low production cost to meet up the industrial need. Though different investigations on first generation to recent fourth generation solar cells prioritize in crossing up the highest conversion efficiency, new structures are essential to be proposed to obtain a remarkable value [1]. Such process requires effort on choosing proper materials and introduction of a novel design with proper integration of those materials in corresponding layers [2-3].

According to the Shockley–Queisser detailed-balance theory, the efficiency limit of a dual-heterojunction solar cell is 42-46% [4-7]. Recently, Marti et al. have reported in a theory that a three-terminal dual-heterojunction bipolar transistor solar cell can have an efficiency of about 54.7% [8]. All of these theories consider wider band gap (1.54-1.8 eV) top layer followed by a lower band gap (0.6-1.1 eV) bottom layer. However, there is no report of such high efficiency solar cell in practice or even in simulation with dual heterojunction structure. Furthermore, the SQ efficiency limit for the intermediate band solar cell is about 63% that essentially works in two-step photon upconversion [9-10]. On the other hand, the impurity photovoltaic (IPV) which essentially enhance the photocurrent due to the absorption of low energy sub-band photon by two-step photon upconversion theoretically offers an efficiency of 77.2% [11]. However, a simulation result on Si IPV solar cell shows an efficiency of 30.5% [12].

Herein, we demonstrate the design and simulation of a high-efficiency dual-heterojunction solar cell that embeds a relatively wider band gap (1.66 eV), $CuGaSe_2$ (CGS) in the bottom of low band gap (1.1 eV), $Cu(In,Ga)Se_2$ (CIGS) layer. It has been observed that the efficiency of the proposed dual-heterojunction structure can reach the theoretical efficiency limit, at least in simulation as no practical device has yet been developed on this framework, as the wide band gap with longer wavelength absorption and adequate doping forms almost Type-II heterojunction with comparatively lower band gap absorber layer on its top [13].

In thin film photovoltaic devices, CIGS-based thin film solar cell has appeared as one of the promising absorber layer candidates with suitable bandgap for solar spectrum absorption and significant rise in efficiency. It has been found with the experimental efficiency to fall in the range of 19-23% [14-17], while its efficiency in numerical analysis lies in between 22-26% [18-19]. The efficiency of the CIGS based cells has been tried to increase with application of SnS, $MoSe_2$, $SnSe_2$, and $BaSi_2$ in BSF layer [19-23]. On the other hand, CGS is a wide band gap (1.5-2.2 eV) p-type semiconductor that has been employed as an absorber layer in CGS-based solar cells [24-26] and top cell in CGS/CIGS tandem solar cells [27-28]. However, there is hardly any report of the use of CGS as BSF layer in the CIGS-based thin film technologies for solar power generation.

In this work, we devise CGS as BSF layer for the improvement in the efficiency of the CIGS thin film solar cell. Here, CGS also act as bottom cell in the proposed CIGS-based *n*-CdS/p-CIGS/$p^+$-CGS dual-heterojunction solar layout. The performance of the designed solar cell has been calculated on the optimized device structure by SCAPS-1D simulator.

## 2. Proposed cell design and simulation model

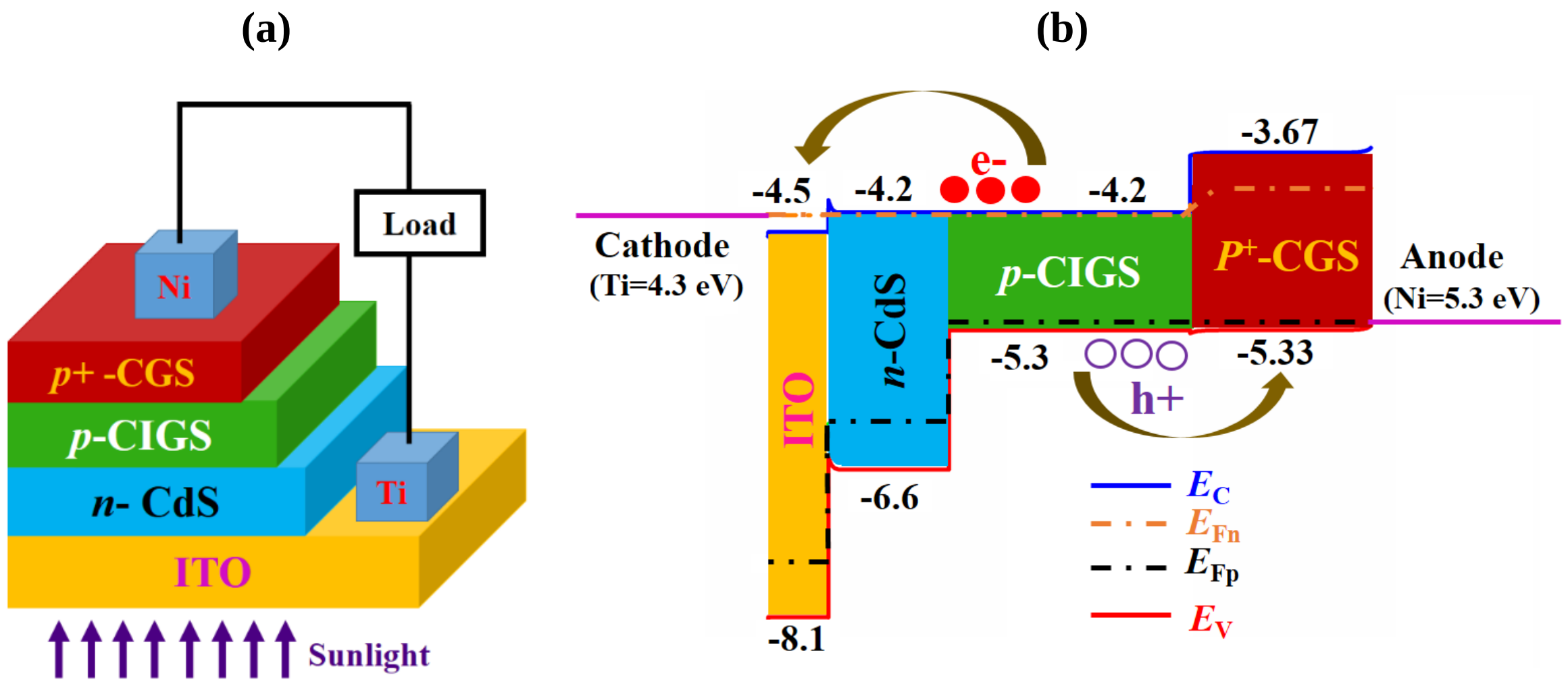

**Fig.1** (a) The schematic structure and (b) illuminated energy band diagram of the modeled *n*-CdS/*p*-CIGS/$p^+$-CGS dual heterojunction solar cell.

The schematic structure and energy band diagram of the modeled *n*-CdS/*p*-CIGS/$p^+$-CGS dual heterojunction solar cell are presented in Fig. 1(a-b), respectively. In this design, light passes through Indium tin oxide (ITO) coated glass substrate and n-type CdS that acts as a window layer. ITO is a popular material for substrate in heterojunction cells due to outstanding transparency [1]. The lights are absorbed by p-type CIGS and then outstretch at $p^+$-type CGS with a band gap of 1.66 eV. Here, p-type CIGS acts as first absorber layer and $p^+$-type CGS plays a dual role as second absorber layer as well as BSF layer. Having respective electron affinity and ionization potential of 3.67 and 5.33 eV for CGS and 4.2 and 5.3 eV for CIGS, CGS can easily form a favorable junction with it.

From Fig. 1 (b) it is evident that quasi-Fermi levels $E_{Fn}$ and $E_{Fp}$ coexist with $E_C$ and $E_V$, respectively in absorber portion. In n-CdS window layer, $E_{Fp}$ lies over $E_V$ whereas $E_{Fn}$ and $E_C$ continue with similar harmony. On the other hand, $E_{Fn}$ has been found to cross over $E_C$ while $E_{Fp}$ and $E_V$ carry on at the same level in $p^+$-CGS BSF layer. Such behavior blocks movement of holes towards the window layer and electrons towards the BSF layer. As a result, back contact can easily collect holes and front contact can easily collect electrons from absorber. With such concept, nickel (Ni) with work function 5.3 eV and titanium (Ti) with work function 4.3 eV have been chosen as back and front contact, respectively.

Our deigned highly efficient solar cell was numerically simulated using a one-dimensional solar cell capacitance simulator (SCAPS-1D) [29] under the illumination of one sun of 100 mW $cm^{-2}$ at global air mass AM 1.5G spectrum and selection of temperature at 300 K. The simulation was accomplished considering ideal values of series and shunt resistances. Gaussian energetic distribution was taken into consideration for single acceptor and donor defects in bulk materials as well as at the absorber interfaces with window and BSF layers. However, radiative recombination coefficient was ignored. To perform the program, absorption coefficient (α) data for $p^+$ CGS BSF layer were collected from literature [30].The similar results were also obtained from absorption data measured in other studies [31-32]. Table 1 shows the parameters for different layers used in this simulation.

**Table 1.** Simulation parameters for *n*-CdS/*p*-CIGS/$p^+$-CGS dual heterojunction solar cell

| **Parameters** | **ITO [33]** | **n-CdS [16]** | **p-CIGS [16]** | **$p^+$- CGS [24,26,34]** |
|---|---|---|---|---|
| Thickness (um) | 0.04 | 0.10 | 0.50 | 0.40 |
| Band gap (eV) | 3.6 | 2.4 | 1.1 | 1.66 |
| Electron affinity (eV) | 4.5 | 4.2 | 4.2 | 3.67 |
| Dielectric permittivity (relative) | 8.9 | 10.0 | 13.6 | 8.15 |
| CB effective density of states ($1/cm^3$) | $2.2\times10^{18}$ | $2.2\times10^{18}$ | $2.2\times10^{18}$ | $2.2\times10^{17}$ |
| VB effective density of states ($1/cm^3$) | $1.8\times10^{19}$ | $1.8\times10^{19}$ | $1.8\times10^{19}$ | $1.8\times10^{18}$ |
| Electron thermal velocity (cm/s) | $1.0\times10^{7}$ | $1.0\times10^{7}$ | $1.0\times10^{7}$ | $1.7\times10^{7}$ |
| Hole thermal velocity (cm/s) | $1.0\times10^{7}$ | $1.0\times10^{7}$ | $1.0\times10^{7}$ | $1.4\times10^{7}$ |
| Electron mobility ($cm^2/Vs$) | 50 | 100 | 100 | 100 |
| Hole mobility ($cm^2/Vs$) | 10 | 25 | 25 | 25 |
| Shallow uniform donor density, $N_D$ ($1/cm^3$) | $1.0\times10^{21}$ | 0 | 0 | 0 |
| Shallow uniform acceptor density, $N_A$ ($1/cm^3$) | $1.0\times10^{7}$ | $1.00\times10^{18}$ | $1.00\times10^{16}$ | $1.00\times10^{17}$ |
| Radiative recombination coefficient ($cm^{-3}/s$) | 0 | 0 | 0 | 0 |
| Defect density ($cm^{-3}$) (above $E_v$ w.r.t. $E_{ref}$ (eV)) | $1.00\times10^{14}$ | $1.00\times10^{14}$ | $1.00\times10^{14}$ | $1.00\times10^{14}$ |

| Defect Type | Single Acceptor | Single Acceptor | Single Donor | Single Donor |
|---|---|---|---|---|
| Capture Cross Section Electrons ($cm^2$) | $1.00\times10^{-15}$ | $1.00\times10^{-15}$ | $1.00\times10^{-15}$ | $1.00\times10^{-15}$ |
| Capture Cross Section Holes ($cm^2$) | $1.00\times10^{-15}$ | $1.00\times10^{-15}$ | $1.00\times10^{-17}$ | $1.00\times10^{-15}$ |
| Energetic distribution | Gaus | Gaus | Gaus | Gaus |
| Reference for defect energy level Et | Above $E_v$ (SCAPS <2.7) | Above $E_v$ (SCAPS <2.7) | Above $E_v$ (SCAPS <2.7) | Above $E_v$ (SCAPS <2.7) |
| Energy level with respect to Reference (eV) | 0.70 | 0.7 | 0.65 | 0.80 |
| Characteristic energy (eV) | 0.10 | 0.10 | 0.10 | 0.10 |

## 3. Results and discussion

### 3.1 Impression of CIGS absorber layer on PV parameters

Fig. 2(a-c) marks out the result of varying thickness, doping concentration and defect densities of *p*-CIGS absorber layer of the proposed *n*-CdS/*p*-CIGS/$p^+$-CGS dual heterojunction solar cell. With the change in thickness from 0.1µm to 0.9 µm, a gradual slight declination in efficiency and fill factor is observed. Open circuit voltage $V_{oc}$ decreases from 1.03 to 0.954 V as well but an opposite increase in short circuit current $J_{sc}$ from 57.85 to 60.73 mA/$cm^2$ occurs at the same period. The increment of short circuit current is obvious as a result of absorbing more photons by thicker layer. On the other hand, the open circuit voltage displays a downturn due to the addendum of associated dark current with thickness enlargement [35].

The fill factor (FF) and photo conversion efficiency (PCE) showed a downward movement from 81.38% to 78.98% and 48.47% to 45.76%, respectively due to negative rate of $V_{oc}$. For further investigations, we consider 0.5 μm thickness as optimized value resulting 47% of PCE with 0.98V of $V_{oc}$ and 59.94 mA/cm$^2$ of $J_{sc}$.

The doping concentration was varied within the range from $1 \times 10^{14}$ to $1 \times 10^{18} cm^{-3}$ which is shown in Fig. 2(b). It is observed that the short circuit current $J_{sc}$ shows almost a constant behavior throughout the scale. However, a sudden drop in efficiency from around 47% to 43% is observed between $1 \times 10^{16}$ to $1 \times 10^{17} cm^{-3}$ of doping level and fill factor follows the similar behavior. The value of open circuit voltage $V_{oc}$ shows almost a constant behavior up to this range and starts to fall above this level. However, with declination in $V_{oc}$, a rise is observed in FF and efficiency as well.

The cell performance has great dependency on bulk defects of CIGS absorber layer which is visualized in Fig. 2(c). $J_{SC}$ have shown almost a constant behavior with slight change from 59.93 to 59.90 mA/cm$^2$. On the contrary, $V_{OC}$ have plummeted from 1.12 to 0.75 V at higher order incorporation of doping. Since photon absorption could be obstructed with higher defects and evolved the lower rate of creating electron-hole pairs (EHPs), the short circuit current is expected in to face a deflation effect. As $J_{SC}$ and $V_{OC}$ both results drop in values, the FF and PCE also exhibit rapid decrement with bulk defects. The FF reduces from 80.56% to 73.81% and PCE is found to downsize at 43.14% in the order of $10^{17}$cm$^{-3}$ of doping concentration.

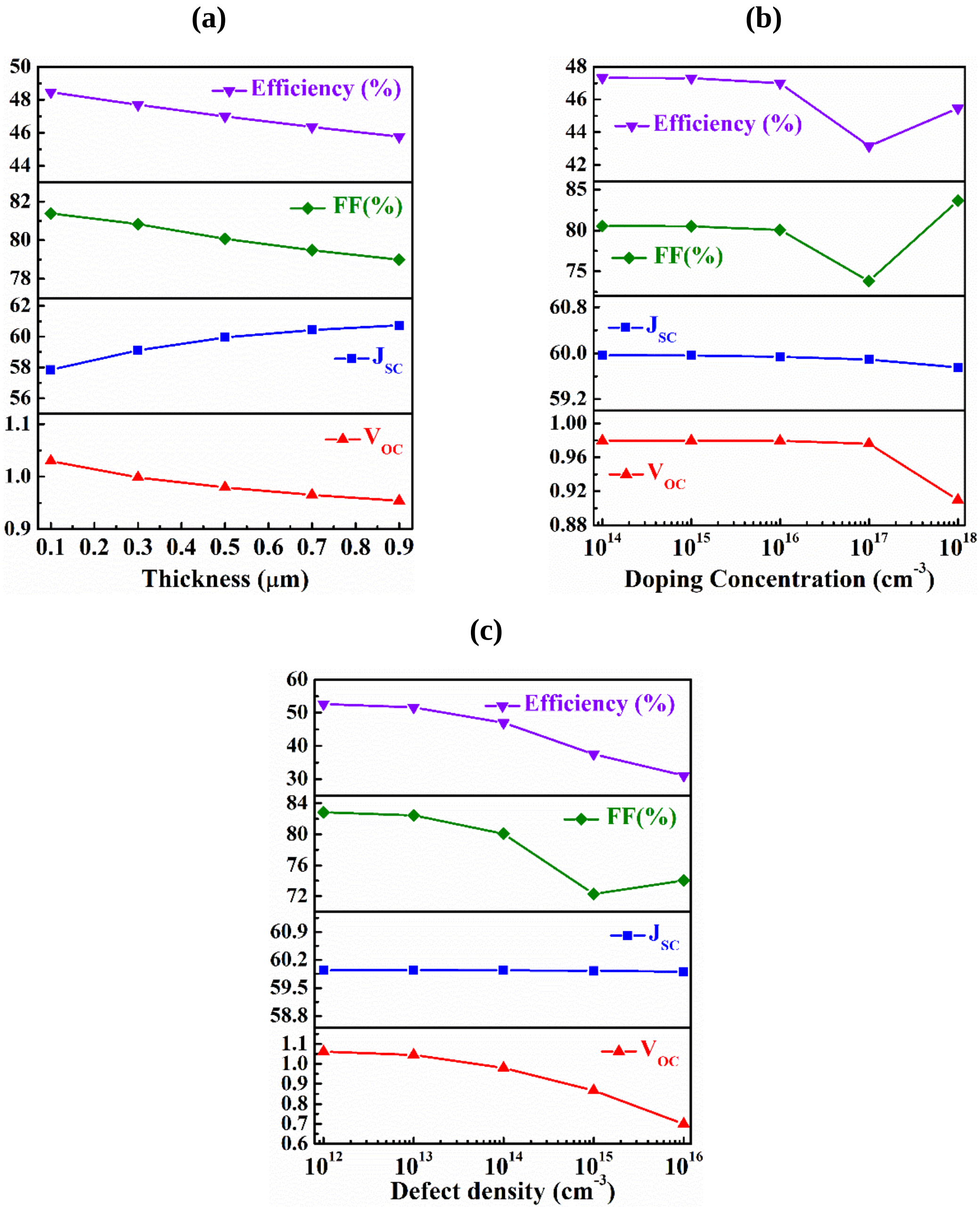

**Fig.2** Simulation findings for variation of (a) thickness (b) doping concentration and (c) defect density in CIGS absorber layer in *n*-CdS/*p*-CIGS/$p^+$-CGS dual heterojunction solar cell.

### 3.2 Impression of window layer on PV parameters

The influence of window layer on PV parameter performance of *n*-CdS/*p*-CIGS/$p^+$-CGS dual heterojunction solar cell has been explored on this step and results are displayed in Fig. 3(a-c). The variation of thickness from a small value of 0.05 to 0.25 µm has not been found to play any significant change on the parameters. The variation of doping concentration was made within $10^{16}$ to $10^{20}$ $cm^{-3}$, short circuit current displayed no change within this range while open circuit voltage resulted with a slight increment and fill factor with a slight decrement after a large amount of doping. However, efficiency seems to increase slightly with amount of doping. The increment in defect density causes a tiny change in efficiency and short circuit close to the order of $10^{16}$ $cm^{-3}$.

(a) (b)

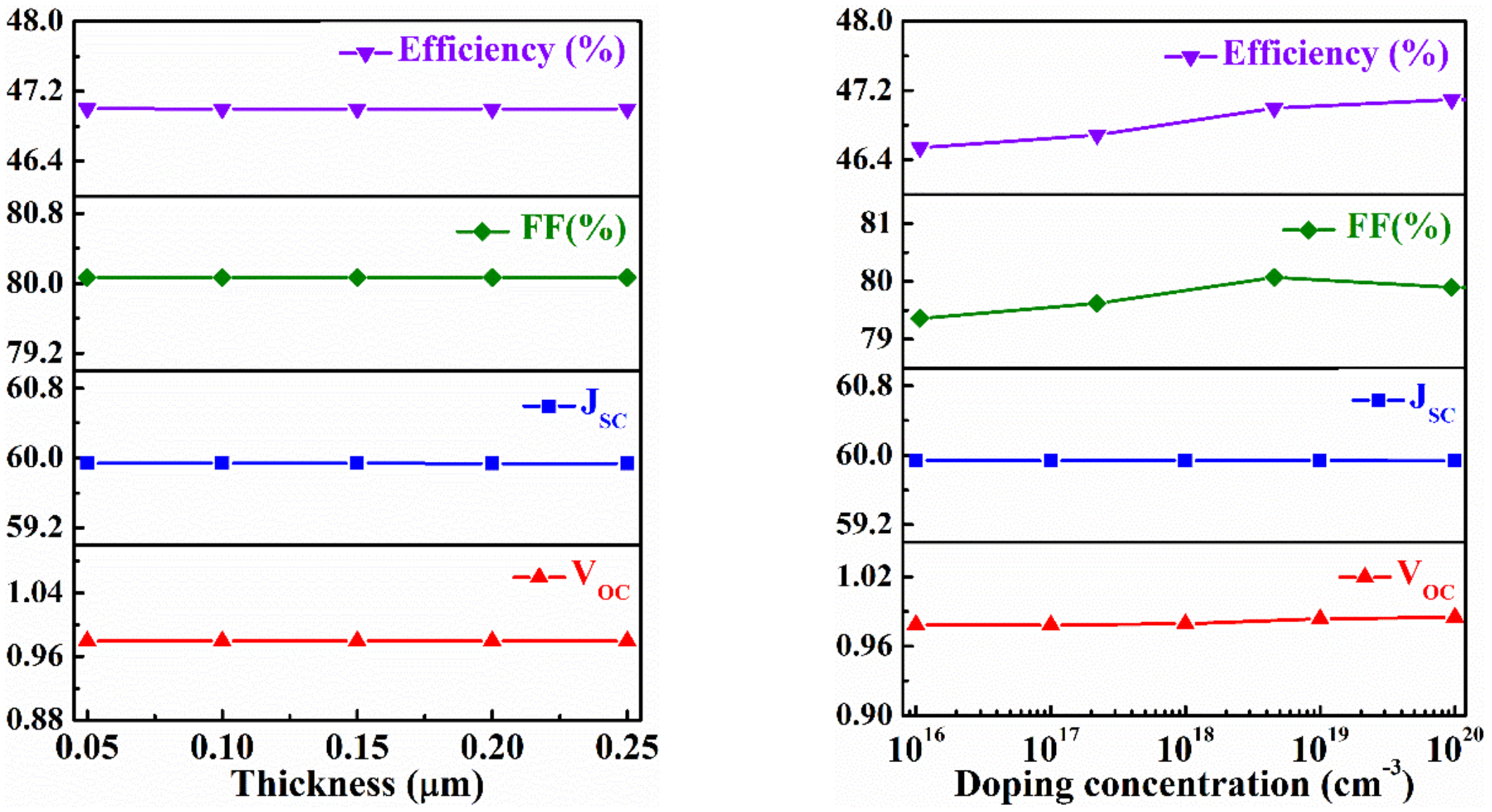

(c)

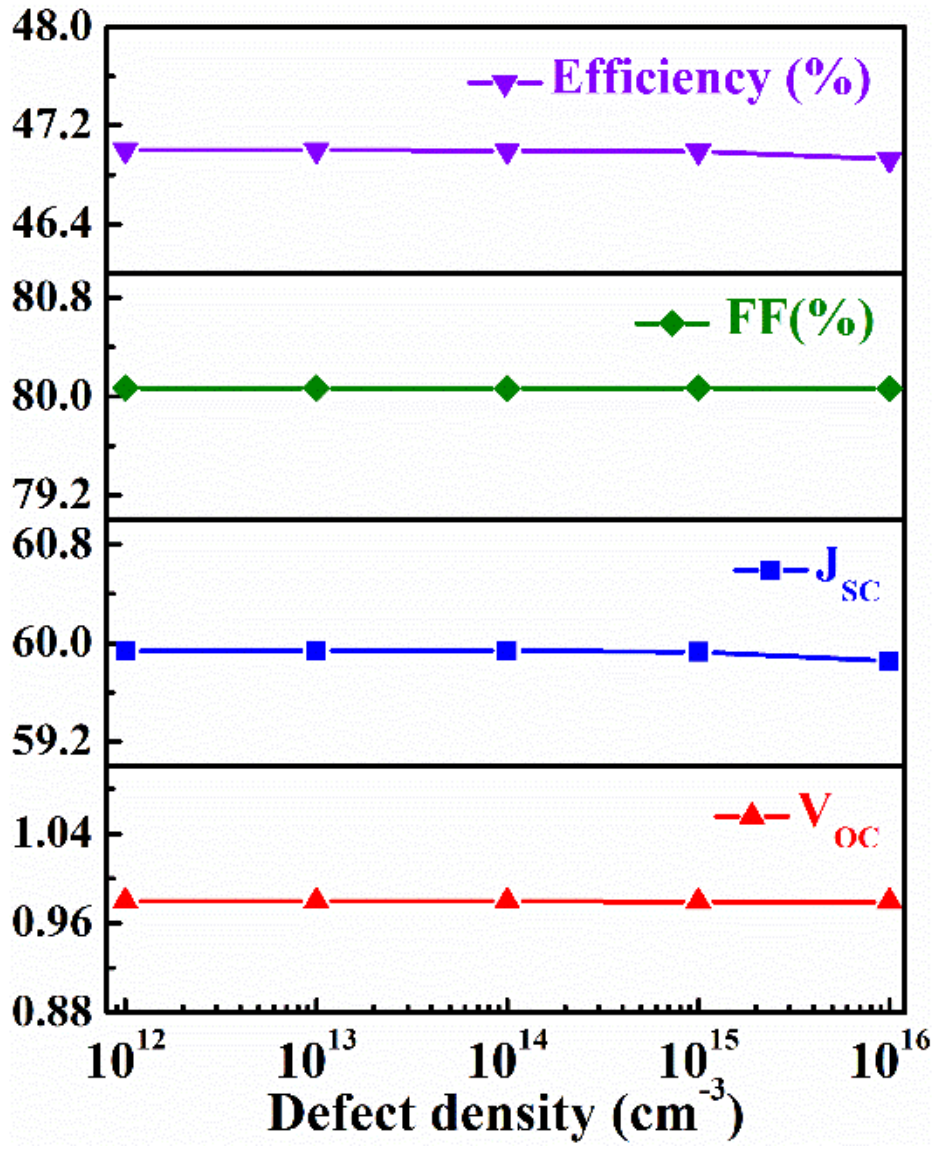

**Fig.3** The impact on PV parameters for the variation of (a) thickness (b) doping concentration and (c) defect density in window layer of *n*-CdS/*p*-CIGS/$p^+$-CGS dual heterojunction solar cell.

### 3.3 Impact of CGS bottom absorber as well as BSF layer on PV parameters

The impact CGS bottom layer along with the effect of absorption of longer wavelength photons in solar spectrum have been studied and ensured during simulation in this section. Figure 4(a-c) presents the obtained results for the variation of CGS layer thickness, doping concentration and defect density, respectively. From figure 4(a), it is seen that the short circuit current went up just for the addition of 0.2 μm on the back of CIGS absorber layer. It continues $J_{sc}$ enhancement with increasing thickness further and 47% of efficiency is obtained at a thickness of 0.4 μm. The combination of high absorption coefficient data, preferred band gap and perfect doping concentration permit absorbing longer wavelength photons in this layer. We have discussed this effect in section 3.4 in details. In Fig. 4(b), it is noticed that the short circuit current sinks with addition of order of doping concentration which in turn playing role in drop in percentage of efficiency [11-13]. With higher doping concentration the parasitic absorption can occur at CGS layer which drains out the cell performance. From Fig.4(c) it can be ensured that the PV parameters are totally independent on bulk defects of thin CGS.

A contour plot for quantum efficiency (QE) as function of thickness and doping concentration at constant photon wavelength of 1100 nm is demonstrated in Fig. 4(d). It is clear from the figure that a climbing in efficiency occurs with increase in thickness up to a certain doping concentration. Here, CGS layer acts as second absorber layer as well as BSF layer, which increases photon absorption with thickness. The maximum QE% has been found up to acceptor concentration level of $10^{18}$ $cm^{-3}$. Above this level QE% is seen to decline with addition of acceptor impurities. Thus, an optimized value could be selected below doping concentration of $10^{18}$ $cm^{-3}$ with CGS layer thickness of 0.4 μm to achieve quantum efficiency above 80%.

(a) (b)

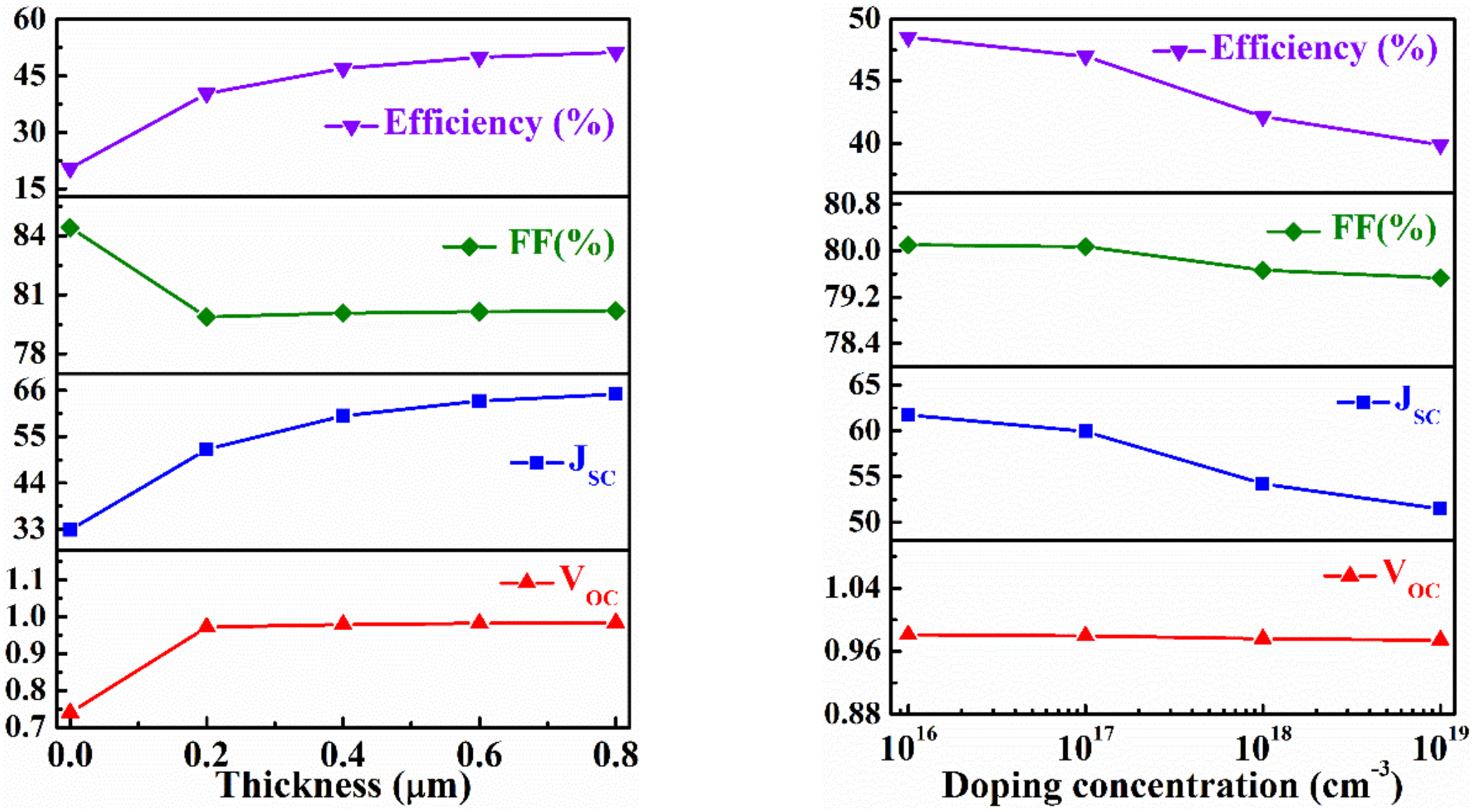

(c) (d)

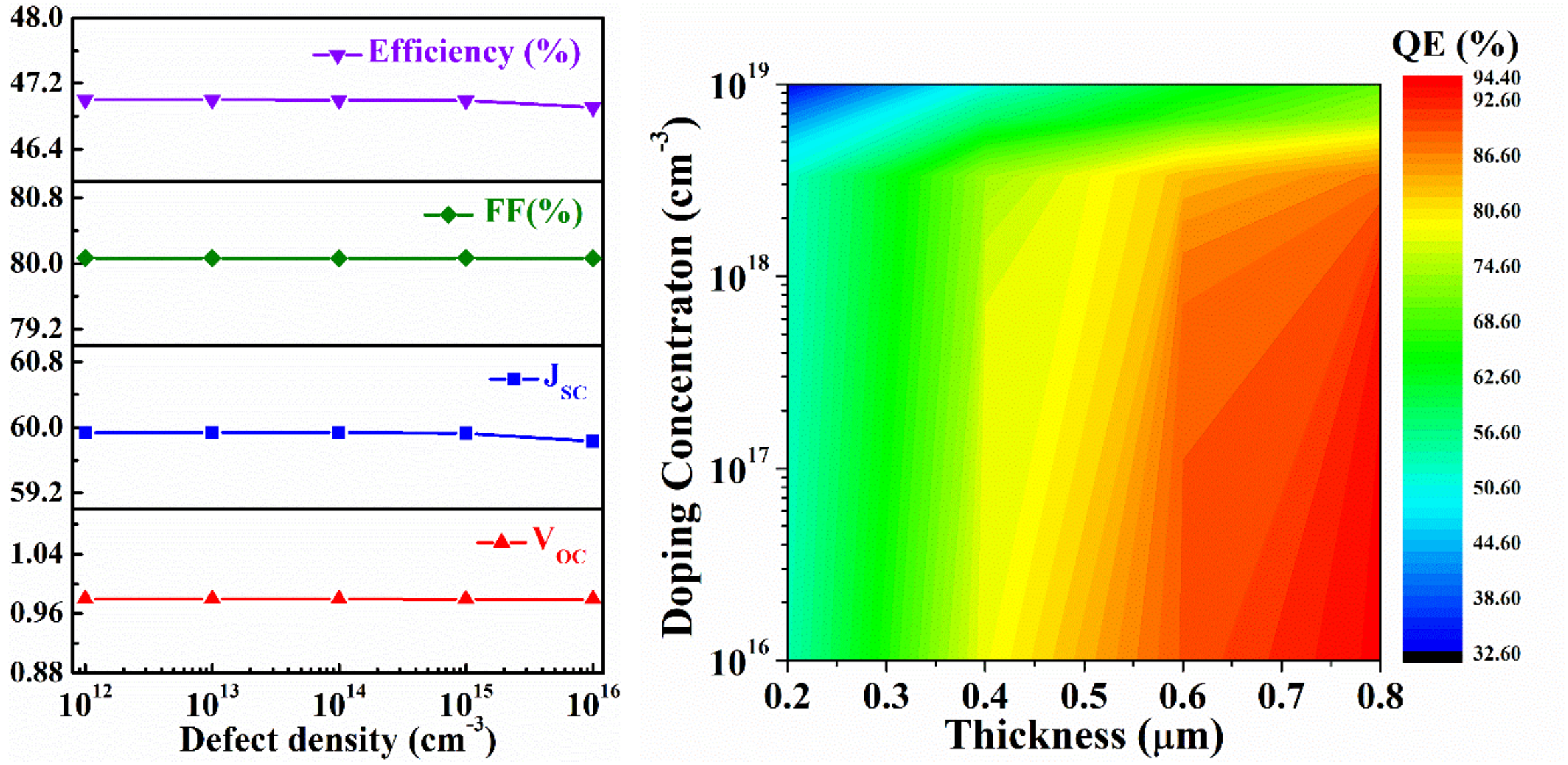

**Fig.4** The impact on PV parameters for variation of (a) thickness (b) doping concentration and (c) defect density and (d) changes in quantum efficiency along with thickness and doping concentration at wavelength of 1100 nm .

### 3.4 Tail-States-Assisted (TSA) two-step photon upconversion

The longer wavelength photon absorption in the second wide band gap absorber layer can be explained considering a two-step photon up-conversion process. The photon absorption and hence contribution to the photocurrent occurs when the materials are adequately doped and it has high absorption co-efficient (α) in the longer wavelength region, i.e. $10^3$ $cm^{-1}$ an wavelength of 1800 nm in this case. If we consider the first and second absorber layers have band gaps of $E_{g1}$ and $E_{g2}$, respectively, the most of the photon with energy $E_p \geq E_{g1}$ will be absorb in the first absorber layer. Then few high energy photon as well as photons with energy $E_p \leq E_{g1}$ will enter the second absorber layer. In convention, these photons will not be absorbed in the second absorber layer as these have energies $E_p < E_{g2}$, even if they are absorbed they will create one free carrier and consequently contribute to the parasitic current [36]. However, these low energy photons may contribute to as a part of two-step photon upconversion process where absorption of two photons in sequence will generate one electron-hole pair and thus contribute to the photocurrent [36-37]. In fact, up-conversion process has been proposed to further improve the efficiency of the solar cell [38-40]. Here, we propose the tail states-assisted (TSA) photon upconversion in dual-heterojunction where upconversion occurs in the wide band gap bottom layer. The band tails states are created close to the conduction or valence band in the heavily doped n or p type semiconductor or when the materials have poor crystallinity [36-37,41]. These states will contribute to the two-step photon upcoversion process.

Fig. 5 shows the mechanism involved in the TSA upconversion process. The figures consider three possible cases in p-type semiconductor to explain the mechanism. It is seen in the figures that a number of states at different levels exists in the band gap of the layer. When a photon with energy $E_p \geq E_{g2}$ enters into the layer it will be immediately absorbed by the layer and create e-h pair. A photon with low energy will be absorbed and generated an electron that will reach the tail state and this electron will absorb a second high energy photon to reach the conduction band, while it has leave a hole in the valance band. Thus an e-h pair is generated.

When a photon with relatively high energy enters the layer it will be absorbed in the layer and an electron will be reached at the top of the tail state leaving a hole in the valance band, where it will absorb a low energy photon to reach the conduction band. This will also generate an e-h pair.

Therefore, by this tail-assisted two-step upconversion process almost all longer wavelength photons that enter the second absorber layer are absorbed. However, the amount of photon absorption will depend on the absorption co-efficient of the layer and the extension of the tail states that may depend on the doping and crystallinity states of the materials.

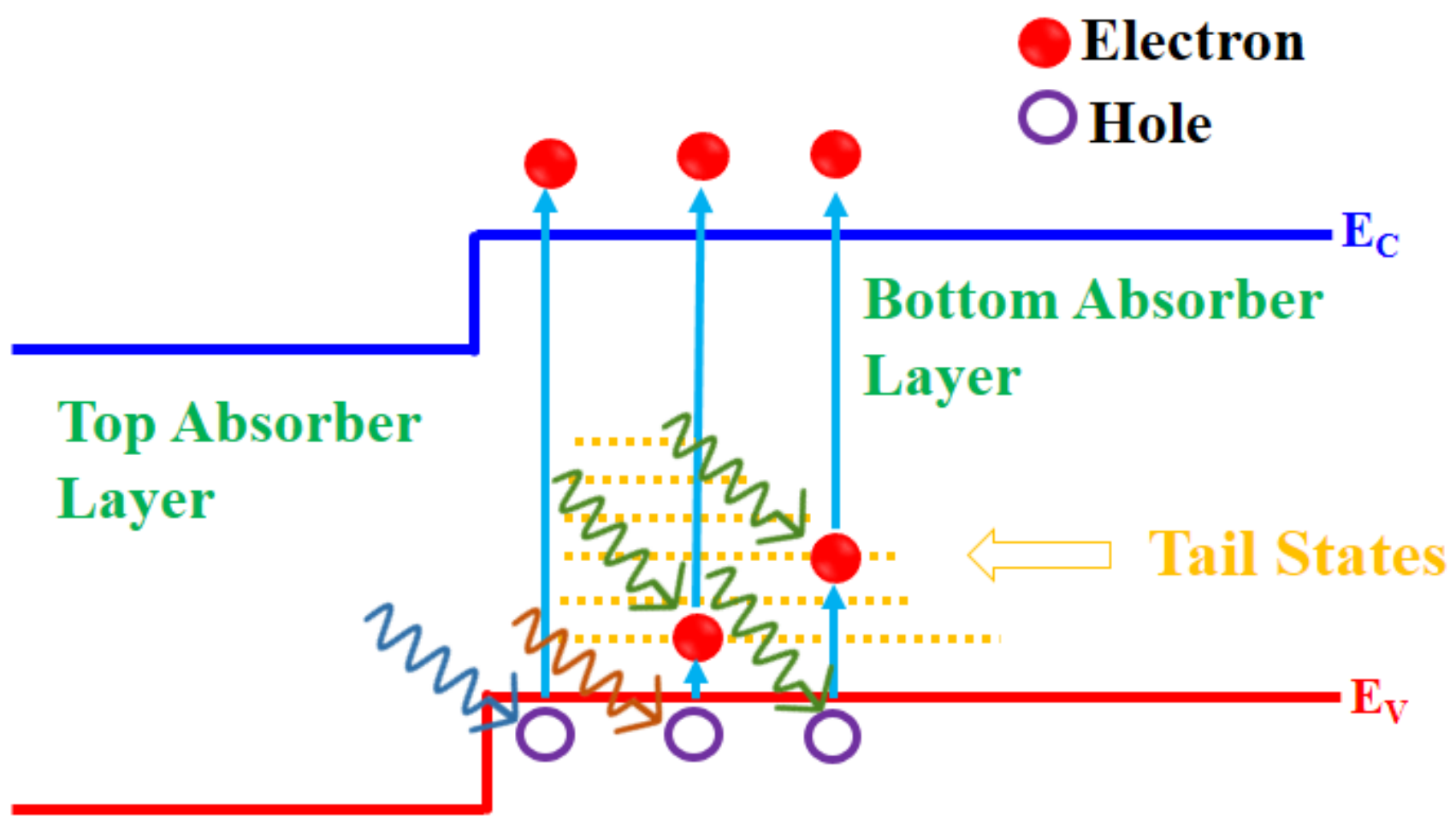

Fig. 5: A simple schematic presentation of the TSA up-conversion process in the bottom layer in the dual-heterojunction solar cell.

### 3.5 Optimized photovoltaic parameters and quantum efficiency

The simulated J-V and QE curves, respectively of *n*-CdS/*p*-CIGS/$p^+$-CGS dual heterojunction solar cell with and without CGS layer are exhibited in Fig. 6(a-b). It is seen that the $J_{SC}$ is enhanced around two times in the dual-heterojunction than in the single-heterojunction due to the addition of CGS layer. The Voc of the DH structure increases by 0.24 V which is mainly originated due to the high built-in potential generated at the CIGS/CGS interface [35,42-43]. In Fig. 6(b), the quantum efficiency for dual-heterojunction is found around 80% in longer wavelength region whereas in single heterojunction the cell i.e. in absence of CGS the structure seems not to perform energy conversion of impinged solar spectra. Such phenomena represent the fact behind increment of photo-conversion efficiency in dual-heterojunction to 47%.

(a) (b)

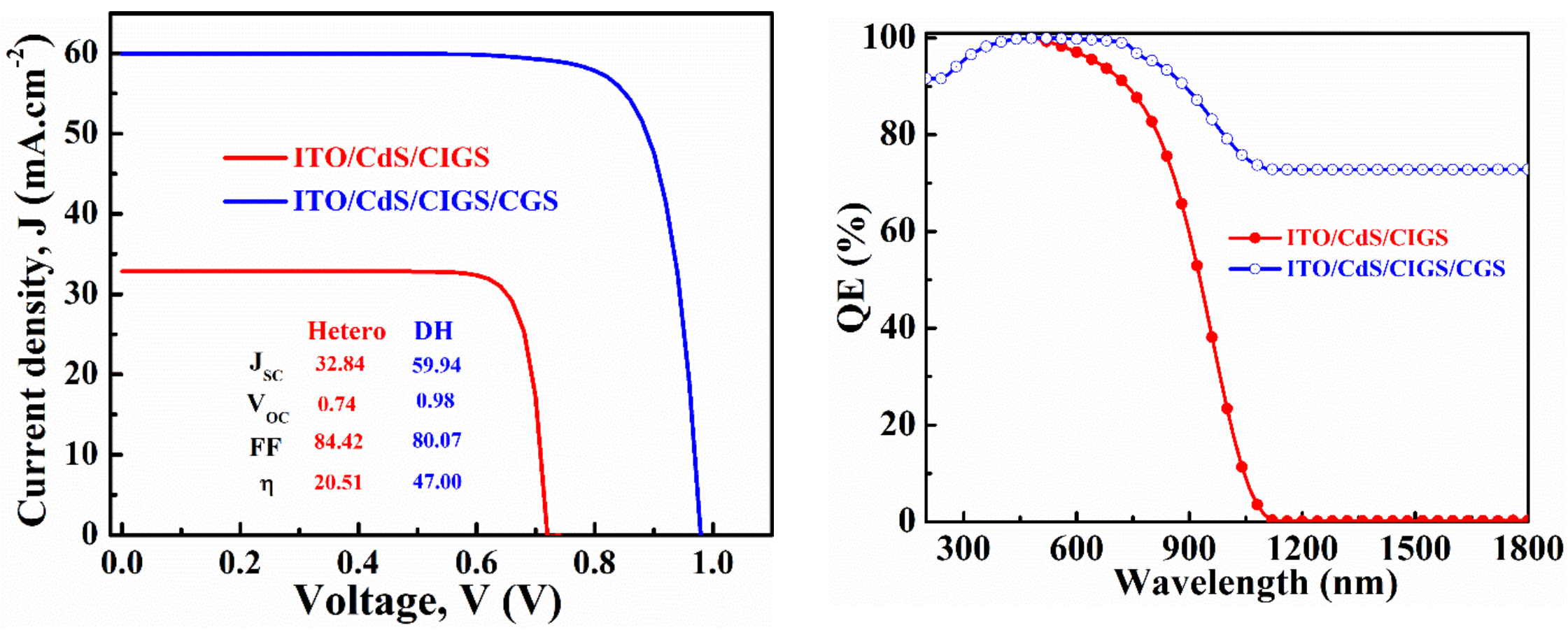

**Fig.6** Simulated J-V and QE cell performance for CIGS-based single- and dual-heterojunction solar cells.

## 4. Conclusion

CIGS-based dual-heterojunction solar cells have been designed and simulated. Based on simulations in SCAPS-1D software, we tried to optimize thickness and doping concentration along with considering bulk and interface defects to reach the SQ limit in dual-heterojunction. The high photocurrent and hence the higher efficiency of the solar cell has been tried to explain using Tail-states-assisted (TSA) two-step photon upconversion that describes the absorption of longer wavelength of solar spectra. The device provides a PCE of 47% with a 0.5 µm thick absorber layer having a doping concentration of $10^{16}$ cm$^{-3}$, and 0.4 µm thick CGS BSF layer having a doping of $10^{18}$ cm$^{-3}$.

## ACKNOWLEDGEMENTS

The authors highly appreciate Dr. Marc Burgelman, University of Gent, Belgium, for providing SCAPS simulation software.

**Corresponding authors:**

*E-mail: jak_apee@ru.ac.bd (Jaker Hossain).

**NOTES:** The authors declare no competing financial interest.